# A Brief Overview of ATM: Protocol Layers, LAN Emulation, and Traffic Management

Kai-Yeung Siu [*]    Raj Jain[†]

**Abstract**

Asynchronous Transfer Mode (ATM) has emerged as the most promising technology in supporting future broadband multimedia communication services. To accelerate the deployment of ATM technology, the ATM Forum, which is a consortium of service providers and equipment vendors in the communication industries, has been created to develop implementation and specification agreements. In this article, we present a brief overview on ATM protocol layers and current progress on LAN Emulation and Traffic Management in the ATM Forum.

## 1 Introduction

The purpose of this article is to introduce the basic aspects of ATM networks for the nonexperts in the area. The length of this short article makes it impossible to cover all important aspects of ATM networks. We shall emphasize some fundamental aspects of ATM that are not addressed in subsequent papers of this special issue. Much of the material in this article is based on existing tutorials on ATM, including [3, 11, 15, 29, 32, 33]. The industrial momentum behind ATM technology and the intensive research interest in ATM has led to a vast and diversified literature in recent years. We have made no attempt to include an exhaustive list of references. Most references we have cited are mainly review articles or documents of ATM Forum. We shall refer those interested in further understanding of the individual topics to the corresponding papers in this special issue and the references therein.

### 1.1 Basic Principles

Future applications are expected to require increasingly higher bandwidth and generate a heterogeneous mix of network traffic. Existing networks cannot provide the transport facilities to efficiently support a diversity of traffic with various service requirements. ATM is potentially capable of supporting all classes of traffic (e.g., voice, video, data) in one transmission and switching fabric technology. It promises to provide greater integration of capabilities and services, increased and more flexible access to the network, and more efficient and economical service.

---

[*]Dept. of Electrical & Computer Engineering, University of California, Irvine, CA 92717. Email: siu@ece.uci.edu
[†]Dept. of Computer and Information Science, The Ohio State University, 2015 Neil Avenue Mall, 497 Dreese Lab, Columbus, OH 43210-1277. Email: Jain@ACM.Org.

ATM carries all traffic on a stream of fixed-size packets (cells), each comprising 5 bytes of header information and a 48-byte information field (payload). The reason for choosing a fixed-size packet is to ensure that the switching and multiplexing function could be carried out quickly and easily. ATM is a connection-oriented technology in the sense that before two systems on the network can communicate, they should inform all intermediate switches about their service requirements and traffic parameters. This is similar to the telephone networks where a fixed path is set up from the calling party to the receiving party. In ATM networks, each connection is called a virtual circuit or virtual channel (VC), because it also allows the capacity of each link to be shared by connections using that link on a demand basis rather than by fixed allocations. The connections allow the network to guarantee the quality of service (QoS) by limiting the number of VCs. Typically, a user declares key service requirements at the time of connection setup, declares the traffic parameters and may agree to control these parameters dynamically as demanded by the network.

## 1.2 The ATM Forum

With the objective of accelerating the convergence of standards and industry cooperation, an international consortium called the ATM Forum was founded to ensure interoperability between public and private ATM implementations and to promote the use of ATM products and services. Although it is not a standards body, the ATM Forum works closely with standards organizations such as the International Telecommunications Union (ITU) and Internet Engineering Task Force (IETF) in developing the definitions for ATM standards. This international consortium has grown from fewer than ten members in 1991 to over 700 members currently, consisting of public and private network equipment vendors and service providers, software companies, as well as government organizations, national research laboratories, and universities.

# 2 ATM Protocol Reference Model

The ATM protocol reference model is based on standards developed by the ITU. Communication from higher layers is adapted to the lower ATM defined layers, which in turn pass the information onto the physical layer for transmission over a selected physical medium. The protocol reference model is divided into three layers: the ATM adaptation layer (AAL), the ATM layer, and the physical layer (Figure 1 [3]).

## 2.1 The ATM Adaptation Layer

The ATM Adaptation Layer (AAL) interfaces the higher layer protocols to the ATM Layer. It relays ATM cells both from the upper layers to the ATM Layer and vice versa. When relaying information received from the higher layers to the ATM Layer, the AAL segments the data into ATM cells. When relaying information received from the ATM Layer to the higher layers, the AAL must take the cells and reassemble the payloads into a format the higher layers can understand. This is called Segmentation and Reassembly (SAR). Four types of AALs were proposed, each supporting a different type of traffic or service expected to be used on ATM networks. The service classes and the corresponding types of AALs were as follows:

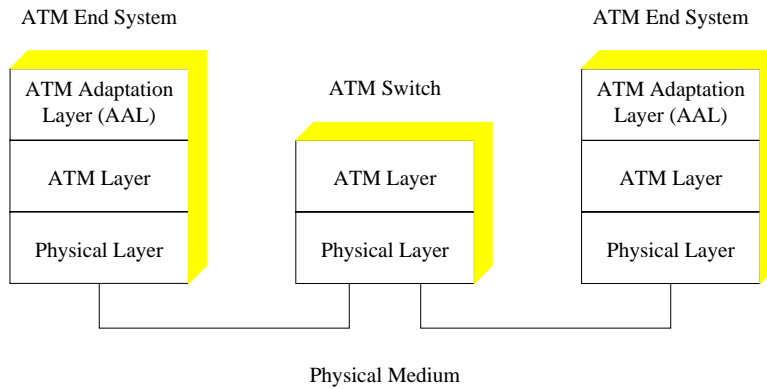

Figure 1: ATM Protocol Structure

- *Class A – Constant Bit Rate (CBR) service*: AAL1 supports a connection-oriented service in which the bit rate is constant. Examples of this service include 64 Kbit/sec voice, fixed-rate uncompressed video and leased lines for private data networks.

- *Class B – Variable Bit Rate (VBR) service*: AAL2 supports a connection-oriented service in which the bit rate is variable but requires a bounded delay for delivery. Examples of this service include compressed packetized voice or video. The requirement on bounded delay for delivery is necessary for the receiver to reconstruct the original uncompressed voice or video. However, at the time of this writing, AAL2 has not been fully developed yet.

- *Class C – Connection-oriented data service*: Examples of this service include connection-oriented file transfer and in general, data network applications where a connection is set up before data is transferred. This service has variable bit rate and does not require bounded delay for delivery. The ITU originally recommended two types of AAL protocols to support this service class, but these two types have been merged into a single type, called AAL3/4. Because of the high complexity of AAL3/4 protocols, the AAL5 protocol has been proposed and is often used to support this class of service.

- *Class D – Connectionless data service*: Examples of this service include datagram traffic and in general, data network applications where no connection is set up before data is transferred. Either AAL3/4 or AAL5 can be used to support this class of service.

Although each AAL is optimized for a specific type of traffic, there is no stipulation in the standards that AALs designed for one class of traffic cannot be used for another. In fact, many vendors of ATM equipments currently manufacture products that use AAL5 to support all the above classes of traffic, and most activities at the ATM Forum have focused on AAL5. The AAL is also important in the internetworking of different networks and services. For more discussion on the issues in AAL5 design, see [28].

In this special issue, the article by Henderson [9] describes a signaling AAL protocol standard recommended by ITU, the Service Specific Connection Oriented Protocol (SSCOP), which incorporates many design principles for a high-speed protocol with lightweight (i.e., reduced processing) operation. The article also contains a list of over thirty references on the topic of AAL, including documents of ITU Recommendation.

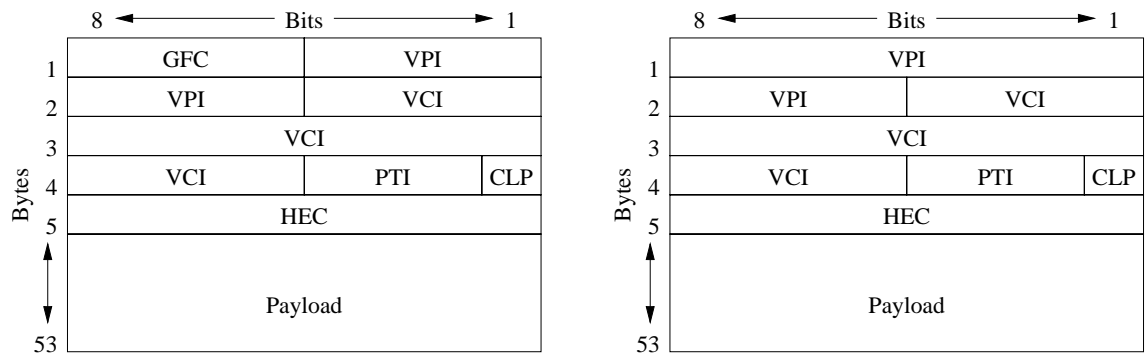

Figure 2: UNI (left) and NNI (right) ATM Cell Format

## 2.2 The ATM Layer

The ATM layer provides an interface between the AAL and the physical layer. This layer is responsible for relaying cells from the AAL to the physical layer for transmission and from the physical layer to the AAL for use at the end systems. When it is inside an end system, the ATM layer receives a stream of cells from the physical layer and transmits either cells with new data or empty cells if there is no data to send. When it is inside a switch, the ATM layer determines where the incoming cells should be forwarded to, resets the corresponding connection identifiers and forwards the cells to the next link. Moreover, it buffers incoming and outgoing cells, and handles various traffic management functions such as cell loss priority marking, congestion indication, and generic flow control access. It also monitors the transmission rate and conformance to the service contract (traffic policing). Traffic management is a hotly debated topic in the ATM Forum, and we shall address the important issues in more details later.

The fields in the ATM header define the functionality of the ATM layer. The format of the header for ATM cells has two different forms, one for use at the user-to-network interface (UNI) and the other for use internal to the network, the network-to-node interface (NNI) (Figure 2). At the UNI, the header dedicates four bits to a function called generic flow control (GFC), which was originally designed to control the amount of traffic entering the network. This allows the UNI to limit the amount of data entering the network during periods of congestion. At the NNI, these four bits are allocated to the virtual path identifier (VPI). Figure 3 gives an illustration of ATM Network Interfaces.

The VPI and the virtual channel identifier (VCI) together form the routing field, which associates each cell with a particular channel or circuit. The VCI is a single-channel identifier; the VPI allows grouping of VCs with different VCIs and allows the group to be switched together as an entity. However, the VPIs and VCIs have significance only on the local link; the contents of the routing field will generally change as the cell traverses from link to link. For the UNI, the routing field contains 24 bits and thus the interface can support over 16 million sessions. At the NNI, the field contains 28 bits, allowing for over 268 million sessions to share a link within a subnet. We refer the readers to the discussion of important issues in Private Network-to-Node Interface

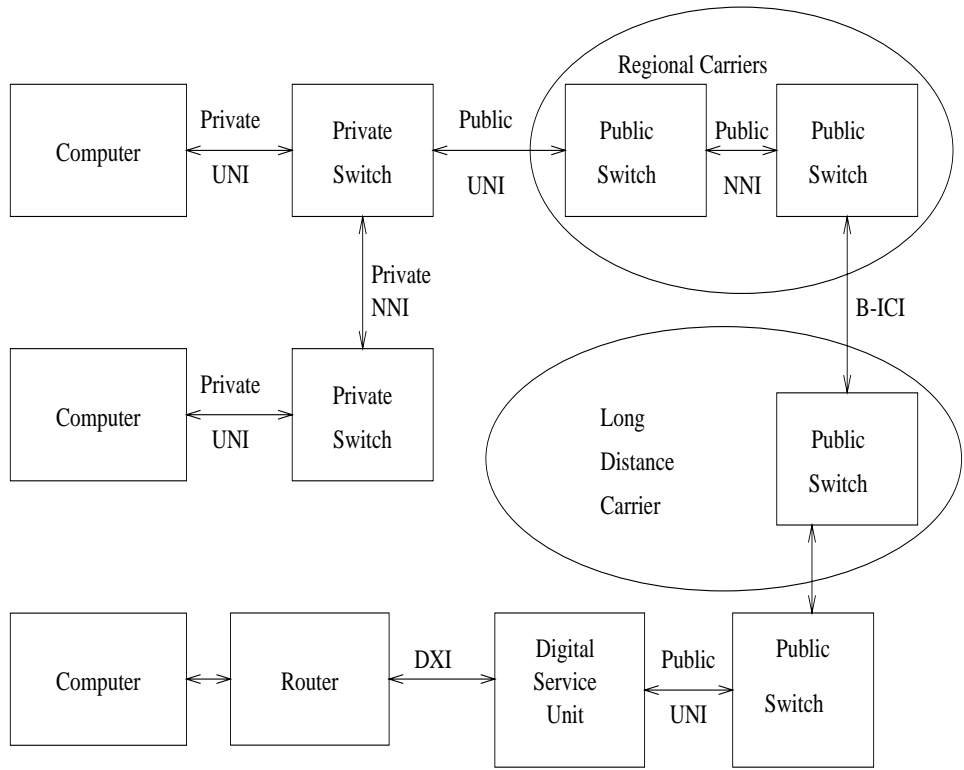

Figure 3: ATM Network Interfaces

(P-NNI) routing to the paper by Lee [18] of this special issue.

The payload type indicator (PTI) field is used to distinguish between cells carrying user data and cells containing control information. This allows control and signaling data to be transmitted on a different subchannel from user data and hence separation of user and control data. A particular combination is used by the AAL if the cell is a part of an AAL5 connection. Another combination is used to indicate that the cell has experienced congestion.

The cell loss priority (CLP) bit provides the network with a selective discard capability. This bit could be set by a user to indicate lower-priority cells that could be discarded by the network during periods of congestion. For example, whereas data applications generally cannot suffer any cell loss without the need for retransmission, voice and video traffic usually can tolerate minor cell loss. One would therefore assign a higher cell loss priority to the CLP bit for voice or video traffic than data traffic. The CLP bit could also be used by the network to indicate cells that exceed the negotiated rate limit of a user.

The header error check (HEC) field is used to reduce errors in the header that cause a misrouting of the cell for one user into another user's data stream. This field contains the result of an 8-bit CRC checking on the ATM header (but not on the data). When a switch or an end system terminates the header, multiple-bit errors will be detected with a high probability. Moreover, a single-bit error can be corrected. This is a desirable since ATM is intended for use on fiber optics link, where the error rate is less than $10^{-9}$ with current modulation techniques. Therefore, single-bit error correction is quite effective in removing most header errors.

## 2.3 The Physical Layer

The physical layer defines the bit timing and other characteristics for encoding and decoding the data into suitable electrical/optical waveforms for transmission and reception on the specific physical media used. In addition, it also provides cell delineation function, header error check (HEC) generation and processing, performance monitoring, and payload rate matching of the different transport formats used at this layer.

The Synchronous Optical Network (SONET), a synchronous transmission structure, is often used for framing and synchronization at the physical layer. In addition to the optical media and line rates defined for SONET, the ATM Forum has proposed a variety of physical layer standards, such as ATM over twisted-pair wire. This will accelerate the acceptance of ATM as a desktop connection technology since existing cabling plants can be retained and the cost per connection will be reduced. We refer the readers to the paper by Rao and Hatamian [21] in this special issue for a discussion on the work by the ATM Physical Layer subworking group at the ATM Forum.

# 3 LAN Emulation

Before ATM fulfills its promise as a cost-effective technology for supporting future broadband multimedia services, the huge legacy of existing LAN applications needs to be readily migrated to the ATM environment. In other words, ATM needs to provide adequate interworking with legacy LANs.

Various approaches have been proposed to support existing LAN applications in ATM networks. One approach is to consider ATM as a new link layer and to modify the existing network layer protocols to this new technology. An example of this approach is the current work taken by the IETF and ATM Forum on specifying mechanisms for running IP and other network layer protocols over ATM. At the time of this writing, the ATM Forum's work on Multiprotocol over ATM is still under development. This approach and the technical challenges will be discussed in the article of Armitage [1] in this special issue.

Another approach is the provision of an ATM protocol to emulate existing LAN services, allowing network layer protocols to operate as if they are still connected to a conventional LAN. The LAN emulation specification defines how an ATM network can emulate a sufficient set of the medium access control (MAC) services of existing LAN technology (e.g., Ethernet and Token Ring), so that higher layer protocols can be used without modification. Unlike the previous approach, which requires ATM users to change their network operating software, such a LAN emulation service can provide a huge cost benefit. The drawback of this approach, however, is that it also prevents higher layer applications from accessing ATM's unique services.

LAN emulation service would be implemented as device drivers below the network layer in ATM-to-legacy LAN bridges and ATM end systems. In an ATM end system adapter, LAN emulation device drivers would interface with widely accepted deriver specifications, such as Network Driver Interface Specification (NDIS) and Open Datalink Interface (ODI) used by TCP/IP and IPX. Figure 4 illustrates the protocol layers for LAN emulation.

In the ATM Forum, the specification task of LAN emulation service was originally taken by the Service Aspects and Applications (SAA) subworking group in early 1993. A new subworking group "LAN Emulation" (LANE SWG) was created later in November 1993 to accelerate the specification work, and in March 1995, the specification was adopted by the ATM Forum.

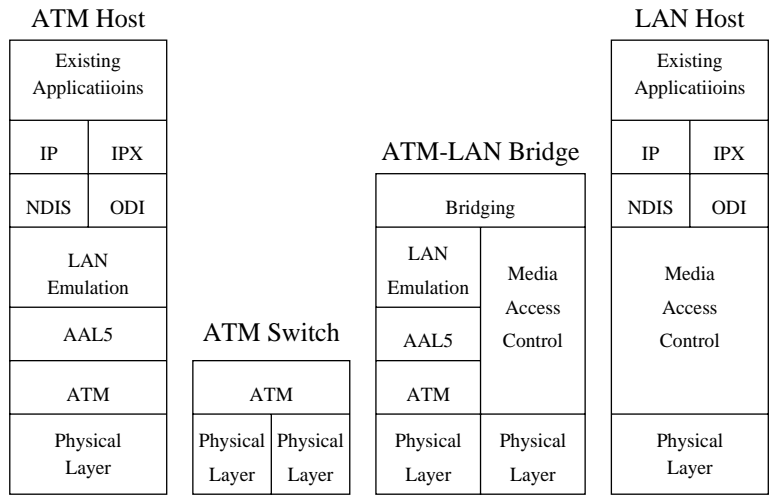

NDIS = Network Driver Interface Specification
ODI = Open Datalink Interface

Figure 4: Protocol Layers for LAN Emulation

An important issue related to the LAN emulation service is bandwidth management. Since existing LANs mainly support "best-effort" service, the LAN emulation service also needs to provide a similar capability to its users. This capability is currently supported by the "available bit rate" (ABR) service. ABR service is important not only for LAN emulation, but also for applications involving bursty data traffic where the bandwidth requirements cannot be predicted. This will be discussed in more details in the section on Traffic Management.

## 3.1 LAN Emulation for Connectionless Service

A major difference between existing LANs and ATM networks is that LANs are connectionless, whereas ATM natively supports only connection-oriented services. Thus, an important function for a LAN emulation service is to be able to support a connectionless service over ATM.

Existing LAN services are typically based on a shared medium. When a source system sends a data frame to a destination, it simply adds to the frame the destination address and broadcasts it to the network. A receiving system will then accept the frame when the destination address included in the frame matches its own address. In a network with multiple LAN segments, bridges and routers are used to handle the forwarding of the frame to the segment to which the destination system is attached.

In a connection-oriented network such as ATM, however, a source system needs to first set up a connection to the destination before it can transfer data frames. This requires the source system to exchange control information with the network using a signaling protocol. The paper by Stiller [27] of this special issue presents a survey of UNI signaling protocols for ATM networks. We shall not go into any further details on the signaling aspect of ATM.

It should be noted that LAN emulation is not designed to provide translational bridging between legacy LANs. A bridge or router is still required to internetwork the different legacy LANs. It will not perform all MAC-layer functions either. For example, there will be no need to perform collision detection on an ATM network.

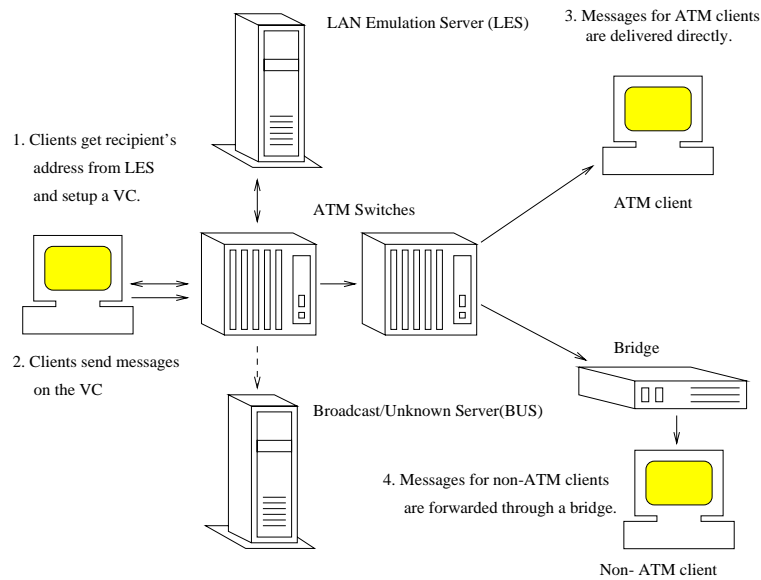

Figure 5: LAN Emulation Architecture

## 3.2 LAN Emulation Architecture

LAN emulation consists of three major components:

- LAN emulation client (LEC) – it is generally located in ATM end systems and serve as a proxy for LAN systems,

- LAN emulation server (LES) – its main function is the support of the LAN Emulation address resolution protocol (LE-ARP), needed by a source LEC to determine the ATM address of the target LEC responsible for a certain destination MAC address.

- Broadcast/Unknown Server (BUS) – its task is to forward all multicast traffic to all attached LECs.

The following describes the message transfer algorithm for a frame with an individual destination MAC address.

1. An LEC maintains an address resolution table that implements a mapping between destination MAC addresses and "data direct" VC. Upon receiving a message from a higher layer, an LEC will first attempt to send the message directly to the destination LEC over a data direct VC.

2. If the destination MAC address is not in the address resolution table, the source LEC sends an LE-ARP request to the LES for the ATM address of the destination.

3. After receiving the LE-ARP reply from the LES containing the target ATM address, the LEC updates its address resolution table. It then establishes a direct VC to that ATM address for transmitting the message. Alternatively, the LEC could send in parallel to the LE-ARP request the initial message with the unknown destination MAC address to the BUS, which will then broadcast the message to all attached LECs.

In the case of a frame with single destination MAC address, the above option of sending messages via the BUS is provided to reduce the message transfer delay caused by the LE-ARP and connection setup procedures and to facilitate the internetworking with stations attached to legacy LANs. In the case of a frame with a multicast MAC address, the source LEC simply sends it to the BUS for broadcasting to all attached stations.

The LAN emulation architecture is illustrated in Figure 5. An in-depth overview of the main architectural issues related to the provision of LAN emulation service over ATM and potential solutions can be found in [29].

# 4 Traffic Management

In order for ATM networks to deliver guaranteed quality of service on demand while maximizing the utilization of available network resources, effective traffic management mechanisms are needed. Almost every aspect of ATM network operation, from signaling requests and routing to network resource allocation and policing, contains some traffic management mechanisms.

## 4.1 Quality of Service Attributes

While setting up a connection on ATM networks, users can specify the following parameters related to the desired quality of service:

- *Peak Cell Rate (PCR)*: The maximum instantaneous rate at which the user will transmit. For bursty traffic, the inter-cell interval and the cell rate varies considerably. The PCR is the inverse of the minimum inter-cell interval.

- *Sustained Cell Rate (SCR)*: This is the average rate as measured over a long time interval.

- *Cell Loss Ratio (CLR)*: The percentage of cells that are lost in the network because of error or congestion and are not delivered to the destination, i.e.,

$$\text{CLR} = \frac{\#\text{ Lost Cells}}{\#\text{ Transmitted Cells}}.$$

Recall that each ATM cell has a cell loss priority (CLP) bit in the header. During periods of congestion, the network will first discard cells with CLP = 1. Since the loss of cells with CLP = 0 is more harmful to the operation of the application, CLR can be specified separately for cells with CLP = 1 and for those with CLP = 0.

- *Cell Transfer Delay (CTD)*: The delay experienced by a cell between network entry and exit points is called the cell transfer delay. It includes propagation delays, queueing delays at various intermediate switches, and service times at queueing points.

- *Cell Delay Variation (CDV)*: This is a measure of variance of CTD. High variation implies larger buffering for delay sensitive traffic such as voice and video.

- *Burst Tolerance (BT)*: This determines the maximum burst size that can be sent at the peak rate. This is the bucket size parameter for the leaky bucket algorithm that is used to control the traffic entering the network. The algorithm consists of putting all arriving cells in a buffer (bucket) which is drained at the sustained cell rate (SCR). The maximum number of back-to-back cells that can be sent at the peak cell rate is called maximum burst size (MBS). BT and MBS are related as follows:

$$\text{BT} = (\text{MBS} - 1)(\frac{1}{\text{SCR}} - \frac{1}{\text{PCR}}) \, .$$

- *Minimum Cell Rate (MCR)*: This is the minimum rate desired by a user.

The first six of the above traffic parameters were originally specified in UNI version 3.0. MCR has been added recently and will appear in the next version of the traffic management document.

## 4.2 Traffic Contract

To provide a guaranteed QoS, a traffic contract is established during connection setup, which contains a connection traffic descriptor and a conformance definition. However, it is not necessary for every ATM virtual connection to have a specified QoS. The reason for this is that if only specified QoS connections are supported by ATM, then a large percentage of the network resources will be wasted. This can happen when one or more connections are not utilizing the full capacity of their QoS contracts. Unspecified QoS contracts can be supported by an ATM network on a "best-effort" basis. Such best-effort services are sufficient for supporting most of the existing data applications.

In general, a traffic contract specifies one of the following classes of traffic:

- *Constant Bit Rate (CBR)*: This class is used for emulating circuit switching, where the bit rate is constant. Cell loss ratio is specified for cells with CLP=0 and may or may not be specified for cells with CLP =1.

- *Variable Bit Rate (VBR)*: This class allows users to send at a variable rate. Statistical multiplexing is used and so there may be small nonzero random loss. Depending upon whether or not the application is sensitive to cell delay variation, this class is subdivided into two categories: real-time VBR (VBR-RT) and nonreal-time VBR (VBR-NRT). While cell transfer delay is specified for both categories. CDV is specified only for real-time VBR. An example of real-time VBR is interactive compressed video while that of nonreal-time VBR is multimedia email.

- *Available Bit Rate (ABR)*: This class is designed for normal data traffic such as file transfer and email. Although the standard does not require the cell transfer delay and cell loss ratio to be guaranteed, it is desirable for switches to minimize the delay and loss as much as possible. Depending upon the congestion state of the network, the source is required to control its rate. The users are allowed to declare a minimum cell rate (MCR), which is guaranteed to the VC by the network. Most VCs will ask for an MCR of zero. Those with higher MCR may be denied connection if sufficient bandwidth is not available.

| Attribute | CBR | VBR-RT | VBR-NRT | ABR | UBR |
|---|---|---|---|---|---|
| CLR for CLP=0 | Specified | | | Specified | Unspecified |
| CLR for CLP=1 | Optional | | | Specified | Unspecified |
| CTD | Specified | Specified | | Unspecified | Unspecified |
| CDV | Specified | Unspecified | | Unspecified | Unspecified |
| SCR and BT | Not applicable | Specified | | Not applicable | |
| PCR and CDVT | Specified | | | Specified | |
| MCR | Not applicable | | | Specified | Not applicable |

Table 1: ATM Layer Service Categories

- *Unspecified Bit Rate (UBR)*: This class is designed for those data applications that want to use any left-over capacity and are not sensitive to cell loss or delay. Such connections are not rejected on the basis of bandwidth shortage (i.e., no connection admission control) and not policed for their usage behavior. During congestion, the cells are lost but the sources are not expected to reduce their cell rate. Instead, these applications may have their own higher-level cell loss recovery and retransmission mechanisms. Examples of applications that use this service are email and file transfer. Of course, these same applications can use the ABR service, if desired.

ABR or UBR are usually specified in the traffic contract when the ATM network is providing a best-effort service. Thus, these two classes of traffic are sometimes referred to as best-effort traffic. The QoS parameters for the above classes of traffic are summarized in Table 1.

### 4.3 Congestion Control Techniques

Congestion control lies at the heart of the general problem of traffic management for ATM networks. In general, congestion arises when the incoming traffic to a specific link is more than the outgoing link capacity. The primary function of congestion control is to ensure good throughput and delay performance while maintaining a fair allocation of network resources to the users. For unspecified QoS traffic such as ABR service, whose traffic patterns are often highly bursty and unpredictable, congestion control poses more challenges than for other services.

As described in [12], one way to classify congestion control schemes is based on the layer of ISO/OSI reference model at which the scheme operates. For example, there are data link, routing, and transport layer congestion control schemes. Typically, a combination of such schemes is used. The selection depends upon the severity and duration of congestion. Figure 6 shows how the duration of congestion affects the choice of the method.

One method to avoid network congestion is to accept a new ATM connection during connection setup phase only when sufficient network resources are available to provide the acceptable QoS. This is called connection admission control (CAC), which is needed for connections where the QoS must be guaranteed. The "busy" tone on telephone networks is an example of CAC. Mechanisms for CAC are currently not standardized and are at the discretion of the network operators.

In addition to CAC, UNI 3.0 also allows traffic shaping using a generic cell rate algorithm (GCRA) and binary (EFCI) feedback congestion control. These mechanisms are described next.

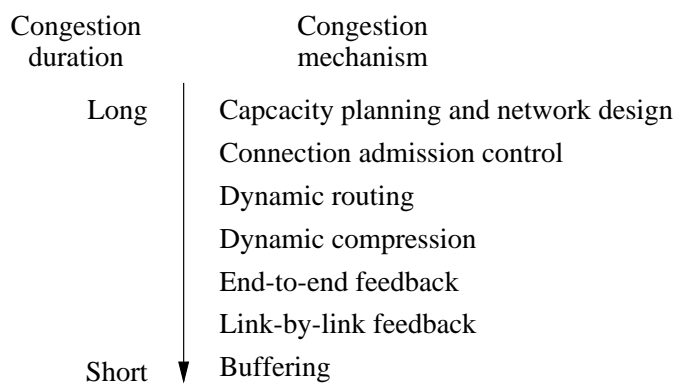

Figure 6: Congestion Techniques for Various Congestion Durations

### Generic Cell Rate Algorithm (GCRA)

The GCRA is also called the "leaky bucket" algorithm, which converts a bursty stream into a more regular pattern. This algorithm essentially works by putting all arriving cells into a bucket, which is drained at the sustained cell rate. If too many cells arrive at once, the bucket may overflow. The overflowing cells are called non-conforming and may or may not be admitted into the network. If admitted, the cell loss priority (CLP) bit of the non-conforming cells may be set so that they will be first discarded in case of overload.

The leaky bucket algorithm is often used by the network to ensure that the input meets the pre-negotiated parameters such as the sustained and peak cell rates. Such "traffic shaping" algorithms are open loop in the sense that the parameters cannot be changed dynamically if congestion is detected after negotiation. In a closed loop (feedback) scheme, however, sources are informed dynamically about the congestion state of the network and are asked to increase or decrease their input rate.

### Feedback Congestion Control

As described earlier (Figure 2), four bits of the cell header at the user-network interface (UNI) are reserved for generic flow control (GFC). It was expected that the GFC bits will be used by the network for flow control; the specific GFC algorithms was left for future definition. However, this approach has been abandoned.

Congestion control techniques, in particular for ABR service, was under intensive debate in the ATM Forum during the past year. An effective congestion control scheme must satisfy several key criteria. In addition to being able to maximally utilize available bandwidth, a good scheme must also provide fairness of network resources to users. Moreover, it must be scalable to a large number of nodes and links with various capacities, robust against slight mistuning of parameters and loss of control cells, as well as low in switch complexity and buffer requirement.

Two major approaches – credit-based and rate-based – have been proposed in the ATM. The credit-based schemes [16, 17] require link-by-link flow control and a separate buffer for each VC (per-VC queueing), which are considered by many switch vendors to be too expensive and inflexible. The rate-based schemes, originally proposed in [19], make use of the "explicit forward congestion indication (EFCI)," a particular combination of the PTI field in the cell header that can be set by the switches during congestion. These rate-based schemes are based on end-to-end control and do not require per-VC queueing nor accounting. Thus, most switch vendors find the rate-based approaches appealing because of their simplicity and implementation flexibility. However, the early rate-based proposals were found to suffer problems of fairness [4]. One approach to fixing the problems is by keeping track of the rate for each VC (per-VC accounting) [7], though such an approach also increases the switch complexity. Many of the current rate-based proposals [2, 10, 13, 14, 22, 26] use an alternative technique called "intelligent marking," originally proposed in [24] (see also [25]), which achieves fairness without the need of per-VC queueing nor per-VC accounting.

An approach that integrates the desirable features of the credit-based proposals and the rate-based schemes was proposed in [30], but it was not accepted by the ATM Forum.

In the September 1994 meeting, the ATM forum voted for rate-based congestion control for supporting ABR service, but without committing to the details of any particular algorithm. We shall not go into the details of the various existing feedback congestion control algorithms. The rate-based congestion control approach and its development at the ATM Forum is described in more details in the article by Ohsaki *et al.* [20] of this special issue. Other sources of reference include the review papers of [15, 6].

## 5 Switch Architecture

Perhaps the most developed aspect of ATM is the switch architecture. Over the past decade, a vast amount of research efforts have been made on studying and designing ATM switches. The field has now become a mature research area and a number of tutorial articles have appeared in the literature. The design of ATM switch architectures is at the discretion of switch vendors. The article by Simcoe and Pei [23] of this issue presents the basic principles of switch design and examines the influence of traffic patterns on the design methodologies.

## 6 Concluding Remarks

In this brief article, we have discussed several key aspects of ATM. We shall conclude by mentioning some other important aspects that are not addressed by the papers in this special issue.

The network-to-node interface (NNI) originally described the interface between two public ATM switches. It is now categorized into Private NNI (P-NNI) and Public NNI. P-NNI describes the ATM switch-to-switch interfaces on a customer's premise. Public NNI, which is also known as the Inter-Switching-System-Interface (ISSI), describes the interface between public ATM switches. ISSI is further categorized into the Intra-LATA ISSI and Inter-LATA ISSI. In an Intra-LATA ISSI, both connected public switches belong to the same Regional Bell Operating Company (RBOC). Inter-LATA ISSI, which is called Broadband Inter-Carrier Interface (B-ICI), describes the link between the ATM switch of an RBOC and that of an inter-exchange carrier such AT&T, MCI, or Sprint. The ATM Forum has a B-ICI subworking group. Currently, much of the work in this subworking group has been completed. Version 1.0 [5] is available from the ATM Forum.

Other important aspects of ATM include efficient network management. The ATM Forum has defined an Interim Local Management Interface (ILMI), which uses the Simple Network Management Protocol (SNMP) and an ATM UNI Management Information Base (MIB) to provide a network administrator with status and configuration information. The ILMI supports bidirectional exchange of management information between management entities related to the UNI ATM Layer and physical layer parameters.

In the Service Aspects and Application (SAA) subworking group in the ATM Forum, the specification of a large number of applications and ATM network operations is under development. This includes an ATM Application Program Interface (API), MPEG-II over AAL5 and/or AAL2, Audio Visual Service over ATM, Frame Relay and SMDS Service over ATM, and a Circuit Emulation Service over AAL1.

Those interested in understanding the specifics of basic ATM operations should find the early UNI Specification Version 3.0 [31] useful.

# References

[1] G. J. Armitage. Multicast and Multiprotocol support for ATM based Internets. This Special Issue.

[2] A. W. Barnhart. Use of the EPRCA with Various Switch Mechanisms. *ATM Forum Contribution 94-0898*, September 1994.

[3] D. Benham. ATM in Loral Area Networks: A Tutorial. Hughes LAN Systems, Spring 1994.

[4] J. Bennett and G. T. Des Jardins. Comments on the July PRCA rate control baseline. *ATM Forum Contribution 94-0682*, July 1994.

[5] B-ISDN Intercarrier Interface (B-ICI) Specification, Version 1.0. ATM Forum, May 1994.

[6] F. Bonomi and K. W. Fendick. The rate-based flow control framework for the available bit rate ATM service. *IEEE Network*, 9(2):25-39, March-April 1995.

[7] A. Charny, D. Clark, and R. Jain. Congestion Control with Explicit Rate Indication. *ATM Forum Contribution 94-0692*, July 1994.

[8] M. De Prycker, R. Peschi, and T. Van Landegem. B-ISDN and the OSI protocol reference model. *IEEE Network*, 7(2):10-18, March 1993.

[9] T. R. Henderson. Design principles and performance analysis of SSCOP: a new ATM Adaptation Layer protocol. This Special Issue.

[10] M. Hluchyj et al. Closed-Loop Rate-based Traffic Management. *ATM Forum Contribution 94-0438R2*, July 1993.

[11] R. Jain. Congestion control in computer networks: issues and trends. *IEEE Network*, 4(3):24-30, May 1990.

[12] R. Jain. Myths about Congestion Management in High Speed Networks. *Internetworking: Research and Experience*, Vol. 3, pp. 101-113, 1992.

[13] R. Jain, S. Kalyanaraman, and R. Viswanathan. The OSU scheme for congestion avoidance using explicit rate indication. *ATM Forum Contribution 94-0883*, September 1994.

[14] R. Jain, S. Kalyanaraman, and R. Viswanathan. Simulation Results: The EPRCA+ Scheme. *ATM Forum Contribution 94-0988*, October 1994.

[15] R. Jain. Congestion Control and Traffic Management in ATM Networks: Recent Advances and A Survey. To appear in *Computer Networks and ISDN Systems*.

[16] H. T. Kung, T. Blackwell, and A. Chapman. Credit-based flow control for ATM networks: Credit update protocol, adaptive credit allocation, and statistical multiplexing. *Computer Communication Review*, 24(4):101-114, October 1994.

[17] H. T. Kung and R. Morris. Credit-based flow control for ATM networks. *IEEE Network*, 9(2):40-48, March-April 1995.

[18] W. C. Lee. Topology Aggregation for Hierarchical Routing in ATM Networks This Special Issue.

[19] P. Newman and G. Marshall. BECN Congestion Control. *ATM Forum Contribution 94-789R1*, July 1993.

[20] H. Ohsaki, M. Murata, H. Suzuki, C. Ikeda, and H. Miyahara Rate-Based Congestion Control for ATM Networks. This Special Issue.

[21] S. K. Rao and M. Hatamian. The ATM Physical Layer. This Special Issue.

[22] L. Roberts. Enhanced PRCA. *ATM Forum Contribution 94-735R1*, September 1994.

[23] R. J. Simcoe and T.-B. Pei. Perspectives on ATM Switch Architecture and the Influence of Traffic Pattern Assumptions on Switch Design. This Special Issue.

[24] K.-Y. Siu and H.-Y. Tzeng. Adaptive Proportional Rate Control for ABR Service in ATM Networks. Technical Report 94-07-01, Electrical and Computer Engineering, University of California, Irvine, July 1994.

[25] K.-Y. Siu and H.-Y. Tzeng. Intelligent Congestion Control for ABR Service in ATM Networks. *Computer Communication Review*, 24(5):81-106, October 1994.

[26] K.-Y. Siu and H.-Y. Tzeng. Adaptive Proportional Rate Control (APRC) with Intelligent Congestion Indication. *ATM Forum Contribution 94-0888*, September 1994.

[27] B. Stiller. A Survey of UNI Signaling Systems and Protocols for ATM Networks. This Special Issue.

[28] T. Suzuki. ATM Adaptation Layer Protocol. *IEEE Communications Magazine*, 32(4):80-83, April 1994.

[29] H. Truong, W. Ellington, J. Le Boudec, A. Meier, and J. Pace. LAN Emulation on an ATM Network. *IEEE Communications Magazine*, 33(5):70-85, May 1995.

[30] H.-Y. Tzeng and K.-Y. Siu. Enhanced Credit-based Congestion Notification (ECCN) Flow Control for ATM Networks. *ATM Forum Contribution 94-0450*, May 1994.

[31] User-Network Interface Specification, Version 3.0. Prentice-Hall, 1993.

[32] R. J. Vetter. ATM concepts, architectures, and protocols. *Communications of the ACM*, 38(2):30-38, 109, February 1995.

[33] R. Vickers. The development of ATM standards and technology: a retrospective. *IEEE Micro*, 13(6):62-73, December 1993.